\documentclass[a4paper,11pt]{article}
\usepackage{jheppub} 
\usepackage{caption}
\usepackage{subcaption}
\usepackage{lineno}
\usepackage[capitalise]{cleveref}
\usepackage[normalem]{ulem}
\usepackage{amsmath}
\usepackage{dsfont}
\usepackage{placeins}
\usepackage[dvipsnames]{xcolor}
\usepackage{braket,xcolor}
\newcommand{\ph}{\varphi}

\title{{The Early Universe as an Open Quantum System: Complexity and Decoherence}}

\author[a]{Arpan Bhattacharyya,}
\author[b]{Suddhasattwa Brahma,}
\author[c,d]{S. Shajidul Haque,}
\author[c]{Jacob S. Lund,}
\author[e]{and Arpon Paul}
\affiliation[a]{Department of Physics, Indian Institute of Technology, Gandhinagar, \\Gujarat-382355, India}
\affiliation[b]{Higgs Centre for Theoretical Physics, School of Physics and Astronomy, University of Edinburgh, Edinburgh EH9 3FD, UK
}
\affiliation[c]{ Department of Mathematics and Applied Mathematics, University of Cape Town,\\ Cape Town-7700, South Africa }
\affiliation [d]{National Institute for Theoretical and Computational Sciences (NITheCS), Private Bag X1, Matieland, South Africa}
\affiliation[e]{School of Physics and Astronomy, University of Minnesota,\\ Minneapolis, Minnesota 55455, USA}

\emailAdd{abhattacharyya@iitgn.ac.in}
\emailAdd{suddhasattwa.brahma@gmail.com}
\emailAdd{shajid.haque@uct.ac.za}
\emailAdd{lndjac009@myuct.ac.za}
\emailAdd{paul1228@umn.edu}
\abstract{In this work, we extend previous results, demonstrating how complexity in an open quantum system can identify decoherence between two fields, even in the presence of an accelerating background. Using the curved-space Caldeira-Leggett two-field model in de Sitter as our toy model, we discover a distinctive feature in the growth of complexity of purification, providing an alternative diagnostic for studying decoherence when the adiabatic perturbation is coupled to a heavy field. This paper initiates a new pathway to explore the features of quantum complexity in an accelerating background, thereby expanding our understanding of the evolution of primordial cosmological perturbations in the early universe.
}

\begin{document}
\maketitle
\flushbottom
\section{Introduction}
Over the past decade, advances in quantum information theory (QIT) have sparked a profound transformation in our understanding of foundational principles in physics and the broader scientific landscape. The application of QIT tools and techniques is providing insights in various fields including quantum field theory, gravity and cosmology. These applications are not just prompting novel questions, but rather inspiring new avenues of research. Many of these ideas possess the potential to fundamentally reshape our understanding on complex quantum systems, emphasizing the central and indispensable role of information, entanglement and other information-theoretic concepts. In this paper, we will employ this approach for understanding the evolution of the early universe.\\

 The standard paradigm of the early universe, namely \textit{inflation}, is a quasi-de Sitter phase of accelerated expansion during which quantum vacuum fluctuations got stretched to enormous wavelengths to source late-time inhomegeneities. These manifest themselves, for instance, as temperature anisotropies that we observe in the cosmic microwave background today. However, this automatically assumes that as the universe evolved through this initial phase of inflation, primordial cosmological perturbations, which are fundamentally quantum in nature, must \textit{decohere} in order to become classical \cite{Burgess:2022nwu, Brandenberger:1990bx, Kiefer:1998qe, Burgess:2006jn, Nelson:2016kjm, Burgess:2014eoa}.\footnote{Interested readers are referred to \cite{Burgess:2017ytm} and the references therein for more details about the application of effective field theories to analyse cosmological perturbations.} Although this explains why it is necessary to consider the early universe as an Open Quantum System  \cite{Brahma:2022yxu, Shandera:2017qkg, Ning:2023ybc, Cao:2022kjn, Gong:2019yyz, Zarei:2021dpb, Brahma:2021mng, Colas:2022kfu}, it also raises the following question -- what is the environment with which observationally-relevant cosmological perturbations can interact in order to undergo such decoherence? By now, it is well-established that there is an abundance of both UV and IR degrees of freedom with which the adiabatic perturbations must interact, and which can lead to decoherence. In this paper, we consider the leading order term for such an interaction with a heavy field, which is typically necessary for a UV-completion to inflation \cite{Colas2022}.\\

\noindent
 However, this brings us to the more important question. Since decoherence has been studied quite extensively in the inflationary community, what more can we learn from the toy model being studied in this work? This is precisely where we believe quantum computational complexity can play a significant role. If indeed the time evolution of the Universe can be described as a quantum circuit \cite{Bhattacharyya:2020rpy, Bhattacharyya:2020kgu,Bhargava:2020fhl,Lehners_2021, Liu:2021nzx, Gomez:2020xdb,Saha:2022onq}, where each transition between different states can be linked to a quantum complexity\footnote{For a more comprehensive discussion regarding the time evolution of complexity, please refer to \cite{Ali:2018fcz, Alves:2018qfv,Ali:2018aon, Chapman:2018hou,Ali:2019zcj}.}, then the dynamics of complexity can be useful to understand how decoherence works in a more general scenario. In other words, although it might be possible to explicitly study decoherence of light adiabatic perturbation in toy models which are simple, things will typically get a lot murkier in the presence of higher order interactions. Take the example of the well-known toy model that we present in this paper. It is purely Gaussian in nature and thus there is a complete lack of (momentum) mode-coupling between the observable and the heavy modes in it (except the trivial case of ${\bf k},-{\bf k}$). In this case, it is easy to keep track of decoherence as the system can be studied exactly in this case. However, recall that GR is non-linear in nature and thus, for any realistic model-building, we must keep higher-order interaction terms which will lead to additional mixing between the observable and the environment modes.\\

\noindent
 In such realistic models of the inflation, where either there are multiple environment fields or mode-coupling between system and environment degrees of freedom (dofs), it is not easy to observe decoherence (and its characteristic timescale) for the observable mode. This is where complexity will come in handy as a complementary probe for decoherence. As is typical while going towards more complicated systems, it is only reasonable to study as many different measures of decoherence as possible, as it is impossible to tell which of these quantum information theoretic quantities will be the easiest to compute in them. Since how the Universe transitions from a quantum state to a classical one is a central aspect for inflation, any such signal for decoherence can, in turn, be used as a smoking gun for the quantum origin of inflation. At the outset, we want to make it clear that we do not propose any such realistic detectable signature for cosmology. Rather, by using  the {\it curved-space Caldeira-Leggett model} \cite{Colas2022} as a toy model we point out a specific signature in the complexity of the light dof which signals that it has decohered due to the environment. We use this toy model only as a benchmark, and expect such a signal in the complexity evolution for any open quantum system which will identify decoherence in it.\\

\noindent
Previously, the \textit{circuit complexity} \cite{NL1,NL2,NL3} of open quantum systems has been studied in various models, as discussed in \cite{Bhattacharyya:2020iic, Bhattacharyya:2021fii, Bhattacharyya:2022rhm}.\footnote{Interested readers are referred to some of the works done in the context of Nielsen complexity \cite{NL1,NL2,NL3} for quantum field theory and quantum many-body systems \cite{Jefferson,Chapman:2017rqy,Bhattacharyya:2018wym,Caputa:2017yrh,Ali:2018fcz,Bhattacharyya:2018bbv,Hackl:2018ptj,Khan:2018rzm,Camargo:2018eof,Ali:2018aon,Caputa:2018kdj,Guo:2018kzl,Bhattacharyya:2019kvj,Flory:2020eot,Erdmenger:2020sup,Ali:2019zcj,Bhattacharyya:2019txx,Caceres:2019pgf,Bhattacharyya:2020art,Liu_2020,Susskind:2020gnl,Chen:2020nlj,Czech:2017ryf,Chapman:2018hou,Geng:2019yxo,Guo:2020dsi,Couch:2021wsm,Erdmenger:2021wzc,Chagnet:2021uvi,Koch:2021tvp,Banerjee:2022ime,Bhattacharyya:2022ren,Bhattacharyya:2023sjr,Bhattacharyya:2022rhm,Bhattacharyya:2021fii,Bhattacharyya:2020iic,ShashiPRB,casagrande2023complexity,Craps:2023rur,Jaiswal:2021tnt,Jaiswal:2020snm,Haque:2021kdm,Bhattacharya:2022wlp}. This list is by no means exhaustive. Interested readers are referred to these reviews and thesis \cite{Chapman:2021jbh, Bhattacharyya:2021cwf,Katoch:2023etn}, and references therein for more details.} Of particular interest to us is the investigation of decoherence in the flat-space Caldeira-Leggett model, which was examined in \cite{Bhattacharyya:2022rhm}. They conducted a comparison of the time evolution of complexity of purification with other standard tools for detecting decoherence, such as linear entropy and entanglement negativity. In the flat-space case, the saturation of complexity was a key feature. It was demonstrated that for the flat-space scenario, the complexity of purification saturates when a state is completely mixed. The saturation timescale for complexity is of the same order as the saturation timescale for linear entropy and entanglement negativity. \\

\noindent
However, when considering the Caldeira-Leggett model in a de Sitter background and within the context of the early Universe, it is not expected that complexity will reach saturation. 
It has been demonstrated in \cite{Bhattacharyya:2020rpy} that 
complexity continues to grow linearly as long as the squeezing parameter increases during the de Sitter phase. Consequently, our attention will now shift towards identifying alternative distinctive features suitable for assessing decoherence in this scenario. On identifying the change in the rate of growth of complexity as such a feature, whose timescale correlates rather faithfully with that of the saturation of the linear entropy, we will go on to conjecture this to be an identifying feature of decoherence from complexity. Thus, while the overall complexity of the system increases forever, as in the free theory \cite{Bhattacharyya:2020rpy}, we will narrow in on a specific feature in the evolution of the complexity, and its correlation with the saturation of linear entropy (and negativity), that will signal decoherence in this model. We hope to study more realistic non-Gaussian systems in the future to see if this feature continues to be a good measure of decoherence.\\

\noindent
The organization of the paper is as follows: In section \ref{sec:model}, we provide details of the model. In the following two sections \ref{sec:LinEntropy} and \ref{sec:cop}, we offer a brief overview of linear entropy and the complexity of purification. In sections \ref{sec:eovCom} and \ref{sec:6}, we delve into the time evolution and the influence of parameters on COP and linear entropy, respectively. In section \ref{sec:decohenrence}, we illustrate how COP can be valuable in detecting decoherence. Finally, in section \ref{sec:8}, we summarize our primary findings and discuss their implications and potential future directions.
\section{The Model} 
\label{sec:model}
In this section, we will delve into the specifics of our model, known as the {\it curved-space Caldeira-Leggett model}, closely following the framework outlined in \cite{Colas2022}. This model includes two massive scalar fields, $\varphi$ and $\chi$, quadratically coupled to each other and minimally coupled to gravity in a de Sitter background defined by the following metric:

\begin{equation}
ds^2 = a^2(\eta)\left(-d\eta^2 + d\vec{x}^2\right),
\end{equation}
where $\eta$ is the conformal time ($-\infty<\eta<0$) and the scale factor $a(\eta) \simeq -1/(H\eta)$, with the constant Hubble parameter $H$. This, of course, is a good proxy for inflation at leading order approximation.
The action can be written as:
\begin{equation}\label{CLaction}
S = -\int d^4x \, \sqrt{-g} \left[\left(\frac{1}{2}g^{\mu\nu}\partial_\mu \ph \partial_\nu \ph + \frac{1}{2}m^2\ph^2\right)+\left(\frac{1}{2} g^{\mu\nu}\partial_\mu \chi \partial_\nu \chi  + \frac{1}{2}M^2\chi^2 \right) + \lambda^2\ph\chi\right],
\end{equation}
where the masses $m$ and $M$ satisfy, $m<3H/2<M$.  In our analysis, we will assume a mass hierarchy between the fields $\ph$ and $\chi$, i.e., $m \ll M$, effectively making the system mass $m$ approach zero.\footnote{Although in the following expressions, we keep the mass $m$ to be general, in our actual computations of complexity we shall set it to 0.} This is the natural assumption to make since we think of $\ph$ field as modelling the adiabatic perturbation during inflation. Thus, to leading order in slow roll approximations, it can be assumed to be massless. We emphasize that this hierarchy in mass is crucial, allowing the light field to act as the adiabatic perturbation, while the heavy one assumes the role of an entropic direction in field space \cite{Colas2022}. The coupling between the two fields is also chosen to be quadratic, with the interaction parameter $\lambda$ also having the dimension of mass. This is the leading order interaction between the cosmological perturbations which mixes the adiabatic and entropic directions.\\

This compelling toy model, with significant physical relevance, provides an ideal framework for probing decoherence due to its exact solvability. As mentioned earlier, our goal is to examine whether the rise in complexity can be associated with decoherence in this model.\\

To decouple the system we now write down the fields $\ph$ and $\chi$, in terms of the Fourier modes multiplied by an additional scale factor $a(\eta)$ as follows:
\begin{equation}\label{CLmodes}
v_\ph(\eta, \mathbf{k})\equiv a(\eta) \int_{\mathbb{R}^3} \frac{d^3\mathbf{x}}{(2\pi)^{3/2}} \ph(\mathbf{x})e^{-i\mathbf{k}\cdot\mathbf{x}},~~~\text{and}~~~v_\chi(\eta, \mathbf{k})\equiv a(\eta) \int_{\mathbb{R}^3} \frac{d^3\mathbf{x}}{(2\pi)^{3/2}} \chi(\mathbf{x})e^{-i\mathbf{k}\cdot\mathbf{x}}.
\end{equation}
The corresponding conjugate momenta ($p_\ph$ and $p_\chi$ respectively) may be evaluated in terms of $v_\ph$ and $v_\chi$ by inverting the expressions in \cref{CLmodes} and rewriting \cref{CLaction} in terms of the modes:
\begin{align}
p_\ph = v_\ph' - \frac{a'}{a}v_\ph,~~~\text{and}~~~p_\chi = v_\chi' - \frac{a'}{a}v_\chi\,.
\end{align}
Then we perform a Legendre transform to obtain the following Hamiltonian:
\begin{equation}\label{CLHamiltonian}
H = \int_{\mathbb{R}^{3+}} d^3\textbf{k} \ \textbf{z}^{\dag}\textbf{H}(\eta,\textbf{k})\textbf{z},
\end{equation}
where
\begin{equation}
\textbf{z} \equiv \begin{pmatrix}
    v_{\varphi}\\
    p_{\varphi}\\
    v_{\chi}\\
    p_{\chi}
    \end{pmatrix} ,~~~ \textbf{H} \equiv \begin{pmatrix}
    \textbf{H}^{(\ph)} & \textbf{V}\\
    \textbf{V} & \textbf{H}^{(\chi)}\\
    \end{pmatrix},
\end{equation}
with
\begin{equation}
\textbf{H}^{(\ph)} \equiv \begin{pmatrix}
    k^2 + m^2a^2 & \frac{a'}{a}\\
    \frac{a'}{a} & 1\\
    \end{pmatrix},~~~
\textbf{H}^{(\chi)} \equiv \begin{pmatrix}
    k^2 + M^2a^2 & \frac{a'}{a}\\
    \frac{a'}{a} & 1\\
    \end{pmatrix},~~~
\textbf{V} \equiv \begin{pmatrix}
    \lambda^2a^2 & 0\\
    0 & 0\\
    \end{pmatrix}.
\end{equation}
Next we promote these Fourier modes to quantum operators and split them into real and imaginary components as   
\begin{align}\label{canonical quantisation}
\hat{\textbf{z}} = \frac{1}{\sqrt{2}}(\hat{\textbf{z}}^R + i\hat{\textbf{z}}^I)\,,
\end{align}
where \(\hat{\textbf{z}}^R\) and \(\hat{\textbf{z}}^I\) are Hermitian. Using the canonical commutation relation,
\begin{align}
[\hat{v}^s_{\alpha}(\textbf{k}),\hat{p}^{s'}_{\beta}(\textbf{k}')] = i \delta^{(3)}(\textbf{k} - \textbf{k}')\delta_{\alpha\beta}\delta_{s,s'}\,,
\end{align}
where $\alpha,\beta \in \{\ph, \chi\}$ and $s,s'\in \{R, I\}$, \cref{CLHamiltonian} may be rewritten with the real and imaginary components of $\hat{\textbf{z}}$ decoupled:
\begin{align}
    H = \frac{1}{2}\sum_{s = R, I}\int_{\mathbb{R}^{3+}} d^3\textbf{k} \ (\hat{\textbf{z}}^s)^{T}\textbf{H}(\eta,\textbf{k})\hat{\textbf{z}}^s\,.
\end{align}
Since the action is quadratic, we can further de-couple the fields $\ph$ and $\chi$ by diagonalizing the mass matrix\footnote{This is like performing a normal mode decomposition of the system.}, with the mixing angle given by
\begin{align}
\theta = \frac{1}{2} \arctan{\left(\frac{2\lambda^2}{m^2-M^2}\right)}\,,
\end{align}
while the eigenvalues of the mass matrix are:
\begin{equation}
\begin{aligned} \label{A10}
m_l^2 &= \frac{1}{2}\left(m^2+M^2-(M^2-m^2)\sqrt{1+\left(\frac{2\lambda^2}{M^2-m^2}\right)^2}\right),\\
m_h^2 &= \frac{1}{2}\left(m^2+M^2+(M^2-m^2)\sqrt{1+\left(\frac{2\lambda^2}{M^2-m^2}\right)^2}\right)\,.
\end{aligned}
\end{equation}
It is evident from \cref{A10} that $m_l < m < 3H/2 < M < m_h$. This establishes a basis, denoted as the  ``light-heavy" basis (as in \cite{Colas2022}), for $\hat{\textbf{z}}$, with the light and heavy field modes, along with their conjugate momenta, represented by $v_l$, $v_h$, $p_l$, and $p_h$, respectively. It is crucial to observe that the interaction term, quadratic in field space, effectively contributes a term to the mass of both the light and heavy modes. To elaborate, initiating with a massless field would imply the existence of an imaginary mass for the light mode, a scenario that will become relevant later on. These field variables are related to the field variables in the $\ph\chi$-basis by:
\begin{align}
\hat{\textbf{z}}_{\ph\chi} = \begin{pmatrix}
    \cos{(\theta)}\mathds{1}_{(2\times2)} & -\sin{(\theta)}\mathds{1}_{(2\times2)}\\
    \sin{(\theta)}\mathds{1}_{(2\times2)} & \cos{(\theta)}\mathds{1}_{(2\times2)}\\
    \end{pmatrix}\hat{\textbf{z}}_{lh}\,,
\end{align}
where $\mathds{1}_{(2\times2)}$ denotes the $2 \times 2$ identity matrix. The Hamiltonian in the $lh$-basis is given as:
\begin{align}\label{lhHamiltonian}
H = \frac{1}{2}\sum_{s = R, I}\int_{\mathbb{R}^{3+}} d^3\textbf{k} (\hat{\textbf{z}}_{lh}^s)^{T}\textbf{H}^{(lh)}(\eta,\textbf{k})\hat{\textbf{z}}_{lh}^s\,,
\end{align}
where,
\begin{equation}
\textbf{H}^{(lh)} \equiv \begin{pmatrix}
    \textbf{H}^{(l)} & \textbf{0}\\
    \textbf{0} & \textbf{H}^{(h)}\\
    \end{pmatrix}\,.
\end{equation}
Here $\textbf{0}$ denotes the \(2 \times 2\) null matrix and,
\begin{equation}
\textbf{H}^{(l)} \equiv \begin{pmatrix}
    k^2 + m_l^2a^2 & \frac{a'}{a}\\
    \frac{a'}{a} & 1\\
    \end{pmatrix},~~~
\textbf{H}^{(h)} \equiv \begin{pmatrix}
    k^2 + m_h^2a^2 & \frac{a'}{a}\\
    \frac{a'}{a} & 1\\
    \end{pmatrix}.
\end{equation}
As anticipated, the light and heavy modes completely decouple in \cref{lhHamiltonian}, and by expanding out the term inside the integrand, the Hamiltonian for each mode is given by:
\begin{equation}
H_{\alpha} = \frac{1}{2}\int_{\mathbb{R}^{3+}}d^3\textbf{k}\left((k^2 + m_{\alpha}^2a^2)v_{\alpha}^2 + p_{\alpha}^2 + 2\frac{a'}{a}v_{\alpha}p_{\alpha}\right)\,,
\end{equation}
where $\alpha \in \{l,h\}$ and the $s$-dependence ($s \in \{R, I\}$) has been left implicit\footnote{Once again, this is possible since the $\{R, I\}$ modes do not mix with each other and propagate independently.}. The equations of motion for each mode \(v_\alpha\) are then given by:
\begin{equation}\label{equs of motion}
v_{\alpha}'' + \left(k^2 - \frac{\nu_{\alpha}^2 - \frac{1}{4}}{\eta^2}\right) v_\alpha = 0\,,
\end{equation}
where
\begin{align} \label{frequency}
\nu_l = \frac{3}{2}\sqrt{1 - \left(\frac{2m_l}{3H}\right)^2}, ~~~\nu_h = \frac{3i}{2}\sqrt{\left(\frac{2m_h}{3H}\right)^2 - 1}\,.
\end{align}
Equation (\ref{equs of motion}) has the following general solution:
\begin{align}
v_{\alpha} = A_{\alpha}\sqrt{z}J_{\nu_{\alpha}}(z) + B_{\alpha}\sqrt{z}Y_{\nu_{\alpha}}(z)\,,
\end{align}
where $z \equiv -k\eta$, and the integration constants $A_{\alpha}$ and $B_{\alpha}$ may be specified by choosing the Bunch-Davies vacuum \cite{Bunch:1978yq} as an initial condition,
\begin{equation}
v_{\alpha} = \frac{1}{2}\sqrt{\frac{\pi z}{k}}e^{i\frac{\pi}{2}(\nu_{\alpha} + \frac{1}{2})}H^{(1)}_{\nu_{\alpha}}(z)\,,
\end{equation}
where $H^{(1)}_{\nu_{\alpha}}$ is the $\nu_{\alpha}$-order Hankel function of the first kind. The conjugate momenta are given by:
\begin{equation}
p_{\alpha} = -\frac{1}{2}\sqrt{\frac{\pi k}{z}}e^{i\frac{\pi}{2}(\nu_{\alpha} + \frac{1}{2})}\left(\left(\nu_{\alpha} + \frac{3}{2}\right)H^{(1)}_{\nu_{\alpha}}(z) - zH^{(1)}_{\nu_{\alpha} + 1}(z)\right)\,.
\end{equation}
As mentioned earlier, the system is characterised by its covariance matrix, whose components are given by:
\begin{equation}
\Sigma_{ij} = \frac{1}{2}\langle \hat{\textbf{z}}_i\hat{\textbf{z}}_j + \hat{\textbf{z}}_j\hat{\textbf{z}}_i\rangle - \langle\hat{\textbf{z}}_i\rangle\langle\hat{\textbf{z}}_j\rangle\,.
\end{equation}
For the $lh$-basis, the covariance matrix admits a block-diagonal form:
\begin{equation}
\Sigma^{(lh)} = 
\begin{pmatrix}
    \Sigma^{(l)} & \textbf{0}\\
    \textbf{0} & \Sigma^{(h)}\\
\end{pmatrix},
\end{equation}
where
\begin{equation}
\Sigma^{(\alpha)} = \begin{pmatrix}
    |v_{\alpha}|^2 & \text{Re}(v_{\alpha}p_{\alpha}^*)\\
    \text{Re}(v_{\alpha}p_{\alpha}^*) & |p_{\alpha}|^2\\
\end{pmatrix}.
\end{equation}
Thus, in the \(\ph\chi\)-basis, the covariance matrix is:
\begin{eqnarray}
\Sigma^{(\ph\chi)} &=& \begin{pmatrix}
    \cos{(\theta)}\mathds{1}_{(2\times2)} & -\sin{(\theta)}\mathds{1}_{(2\times2)}\\
    \sin{(\theta)}\mathds{1}_{(2\times2)} & \cos{(\theta)}\mathds{1}_{(2\times2)}\\
    \end{pmatrix} \Sigma^{(lh)}
    \begin{pmatrix}
    \cos{(\theta)}\mathds{1}_{(2\times2)} & \sin{(\theta)}\mathds{1}_{(2\times2)}\\
    -\sin{(\theta)}\mathds{1}_{(2\times2)} & \cos{(\theta)}\mathds{1}_{(2\times2)}\\
    \end{pmatrix}\\
\label{covmat} 
&=& \begin{pmatrix}
    \Sigma^{(\ph\ph)} & \Sigma^{(\ph\chi)}\\
    \Sigma^{(\ph\chi)} & \Sigma^{(\chi\chi)}\\
    \end{pmatrix},
\end{eqnarray}
where
\begin{align}\label{eq:covmatrix}
    &\Sigma^{(\ph\ph)} = \cos^2(\theta)\Sigma^{(l)} + \sin^2(\theta)\Sigma^{(h)}\,,\\
    &\Sigma^{(\ph\chi)} = \cos(\theta)\sin(\theta)(\Sigma^{(l)} - \Sigma^{(h)})\,,\\
     &\Sigma^{(\chi\chi)} = \cos^2(\theta)\Sigma^{(h)} + \sin^2(\theta)\Sigma^{(l)}\,.
\end{align}

The role of the time-dependent background is to provide a ``squeezing'' term which creates zero-momentum pairs of particles, while the interaction term leads to entanglement of the two fields. Nevertheless, since the action is quadratic, the Fourier modes of each of these states remain decoupled from each other, and it is sufficient to consider a particular pair of Fourier modes $({\bf k}, -{\bf k})$ to see complexity and decoherence in this model.

\section{Linear Entropy ($S_l$)}\label{sec:LinEntropy}
As mentioned at the very outset, our main goal is to understand decoherence in this model. As the light field interacts with the heavy one through the Guassian interaction term, quantum entanglement builds up. It is well-known that decoherence is exhibited when tracking the quantum state of the light field alone. This is due to the introduction of ``mixedness'' in the field, which characterizes the loss of information in the system. We reiterate that the ease of computing this degree of decoherence so cleanly is only possible in this system due to its Gaussian nature which implies that a given Fourier mode, ${\bf k}$, can only interact with with a very small subset of environment modes $-{\bf k}$. \\

A commonly used measure to quantify the degree of decoherence experienced by subsystem-$A$ due to subsystem-$B$ is called purity. It is defined as follows \cite{Zurek:2003zz,Peters_2004}:
\begin{equation}
\gamma = \text{Tr}[\hat{\rho}_A^2],
\end{equation}
where $\hat{\rho}_A$ is the density matrix of subsystem-$A$. The purity $\gamma$ lies within the range $1/d \leq \gamma \leq 1$, where $d$ represents the dimension of the Hilbert space. A pure state corresponds to $\gamma=1$, while a completely mixed state yields $\gamma=1/d$. Similarly, the linear entropy $S_l$ can be defined as:
\begin{equation}
S_l = 1 - \gamma\,.
\end{equation}
The linear entropy provides an alternative measure of decoherence, with values ranging from $0$ for pure states to $1-1/d$ for maximally mixed states. Since we are studying an interacting, albeit Gaussian, field theory in de Sitter, the Hilbert space is infinite dimension and thus, $S_l \rightarrow 1$ for a fully decohered system.\\

As the state of interest is Gaussian, we may compute the purity $\gamma(\eta)$ from the covariance matrix (\cref{eq:covmatrix}) as outlined in \cite{Colas:2021llj}:
\begin{equation}
    \gamma(\eta) = \frac{1}{4\det(\Sigma^{(\ph\ph)})}\,.
\end{equation}
Thus, the linear entropy \(S_l(\eta) \equiv 1 - \gamma(\eta)\) is given by:
\begin{equation}
    S_l(\eta) = 1 - \frac{1}{4\det(\Sigma^{(\ph\ph)})}\,.
\end{equation}
We will study the evolution of the linear entropy for this system, and correlate with the complexity later on to see if the latter can detect when the system decoheres.
\section{Complexity of Purification} \label{sec:cop}
The mixed state, represented by the density matrix $\hat{\rho}_{\rm mix}$, can be purified to a pure state $\ket{\psi}$ in an enlarged Hilbert space $\mathcal{H}_{\rm pure} \otimes \mathcal{H}_{\rm anc}$, where $\mathcal{H}_{\rm anc}$ represents an ancillary set of degrees of freedom. To ensure that the original mixed state is obtained upon tracing out the ancillary Hilbert space, the trace of the density matrix of $\ket{\psi}$ over $\mathcal{H}_{\rm anc}$ must yield $\hat{\rho}_{\rm mix}$:
\begin{equation}
\rm {Tr}_{\mathcal{H}{\rm anc}}\left(\ket{\psi}\bra{\psi}\right) = \hat{\rho}_{\rm mix}\,.
\end{equation}
The purification process preserves the expectation values of operators acting in the initial Hilbert space $\mathcal{H}$. Thus, observables are conserved under purification:
\begin{equation}
\langle \hat{\mathcal{O}} \rangle = {\rm Tr}_{\mathcal{H}}(\hat{\rho}_{\rm mix}\hat{\mathcal{O}}) = {\rm Tr}_{\mathcal{H}_{\rm anc}}(\bra{\psi}\hat{\mathcal{O}}\ket{\psi})\,.
\end{equation}

Since the choice of ancillary Hilbert space $\mathcal{H}_{\rm anc}$ is not unique, the purification process is arbitrary and dependent on this choice. A set of pure states $\ket{\Psi}_{\alpha,\beta, ...}$ can be constructed, parameterized by $\alpha,\beta, ...$, that satisfy the purification requirement. The purification process can be optimized by minimizing the quantity of interest (such as complexity or entanglement entropy) with respect to these parameters. In this study, we focus on the complexity of the mixed state and aim to find the minimum complexity among the set of purifications $\ket{\Psi}_{\alpha,\beta, ...}$ of $\hat{\rho}_{\rm mix}$, obtaining the complexity of purification (COP):

\begin{equation}
\rm{COP} \equiv \underset{\alpha,\beta, ...}{\text{min}} \mathcal{C}\left(\ket{\Psi}_{\alpha,\beta,...}, \ket{\Psi_R}\right)\,.
\end{equation}
Given that we are dealing with a Gaussian open quantum system, we follow the same approach as \cite{Bhattacharyya:2021fii}. Hence, the complexity of purification (COP) is given by
\begin{align}\label{COP}
    \rm{COP} = \min_{\text{Im}(\beta)}\frac{1}{2}\sqrt{\sum_{i = 1}^2\left(\ln\left(\frac{|\omega_i|}{\omega_0}\right)^2 + \arctan\left(-\frac{\text{Im}(\omega_i)}{\text{Re}(\omega_0)}\right)^2\right)}\,,
\end{align}
where
\begin{eqnarray}
    \omega_1 &\equiv& \frac{1}{2}\left(\alpha + \beta + \sqrt{(\alpha - \beta)^2 + 4\tau^2}\right)\,,\\
    \omega_2 &\equiv& \frac{1}{2}\left(\alpha + \beta - \sqrt{(\alpha - \beta)^2 + 4\tau^2}\right)\,,\\
     \omega_0 &\equiv& 1\,,
\end{eqnarray}
and
\begin{equation}
    \alpha = \Sigma^{(\ph\ph)}_{11}, ~~~ \text{Re}(\beta) = 2\Sigma^{(\ph\ph)}_{22}, ~~~ \tau^2 = 4\left(\Sigma^{(\ph\ph)}_{12}\right)^2 - 1 + 4i\Sigma^{(\ph\ph)}_{12}\,.
\end{equation}
Note that no direct expression is provided for \(\text{Im}(\beta)\). Thus \(\text{Im}(\beta)\) is treated as a free parameter, and in order to compute COP, the expression in \eqref{COP} is minimised over \(\text{Im}(\beta)\).

The reason to focus on this intriguing model is that it presents several notable characteristics. Firstly, it considers an accelerating background (de Sitter) that serves as a time-dependent external source, due to which we have particle creation in the usual sense. Additionally, in this model, the light scalar field, $\varphi$, which is the proxy for scalar adiabatic perturbations, is coupled to another massive scalar field $\chi$, which represents an entropic degree of freedom. This coupling $\lambda$ allows for energy loss over time for the observable sub-system. 
We will investigate its role in the time evolution of COP. 

\section{Time Evolution of Complexity}
\label{sec:eovCom}
The complexity of scalar fields in de Sitter, especially that of light or massless ones, have been shown to increase due to the squeezing associated with the background spacetime \cite{Bhattacharyya:2020rpy, Bhattacharyya:2020kgu}. The reasoning behind this is that one assumes the reference state in this computation to be the Fock vacuum, for short wavelength modes deep inside the horizon, and the target state is the squeezed one once the mode exits the horizon. As the wavelength of the mode gets more and more redshifted, the squeezing parameter increases and thus the complexity rises unboundedly. This, of course, is only true for an idealized de Sitter space and not if one has a subsequent radiation era post inflation.\\ 

\noindent
In the following, we will show the evolution of the complexity with respect to the interaction parameter $\lambda$. In a later section we will compare COP with the linear entropy (and log negativity in Appendix \ref{app:neg}), in order to investigate signatures of decoherence from complexity.\\

\begin{figure}[htb!]
    \centering
    \begin{subfigure}[t]{0.40\textwidth}
        \centering
        \includegraphics[width=\linewidth]{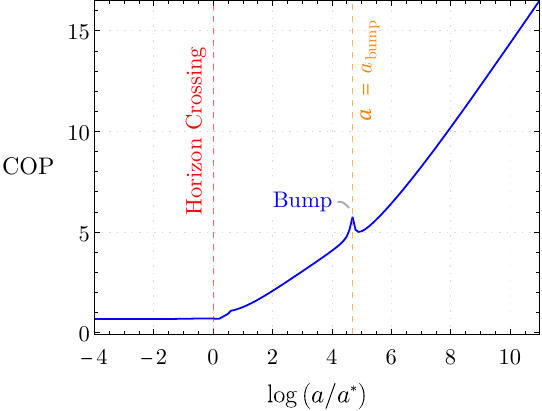} 
        \caption{$\lambda^2 < 3MH/2$} \label{fig:typeI}
    \end{subfigure}
    \begin{subfigure}[t]{0.40\textwidth}
        \centering
        \includegraphics[width=\linewidth]{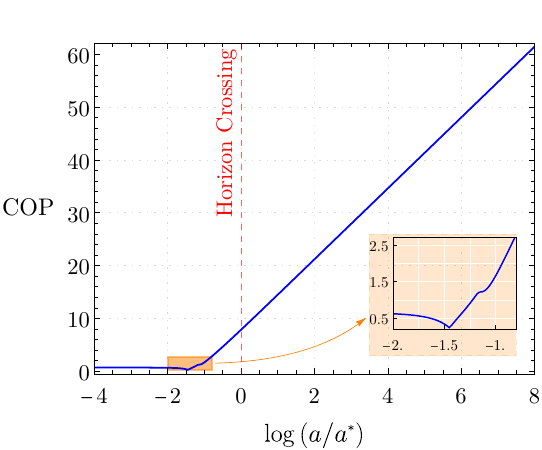} 
        \caption{ $\lambda^2 > 3MH/2$} \label{fig:typeII}
    \end{subfigure}
    \caption{Generic features of the time evolution of COP under different regimes of the coupling strength. Here, we set $m = 0, M = 10, H = 1$ and $k = 1$. The coupling strength $\lambda = 0.4$ and $7,$ in the left and right panel respectively. The red dashed line shows the timescale of horizon crossing, i.e, $a= a^* \equiv k/H$. The orange dashed line in the left panel indicates the timescale of the appearance of the bump, i.e, $a = a_{\text{bump}}$ for regime (i). The inset of the right panel highlights the change of the COP growth-rate with  the appearance of a small bump in regime (ii).}\label{fig:COP_generic}
\end{figure}
\begin{figure}
    \centering
    \begin{subfigure}[t]{0.40\textwidth}
        \centering
\includegraphics[width=\linewidth]{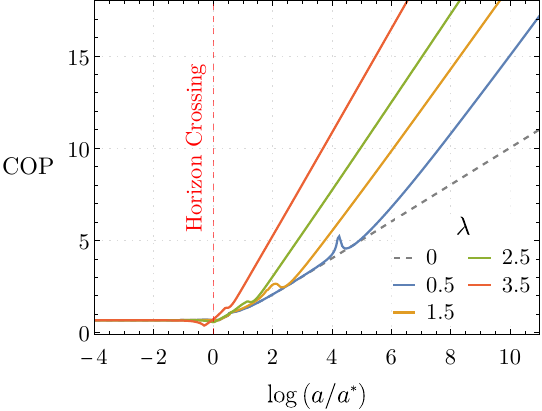} 
        \caption{${\lambda}^2 < 3MH/2$} \label{fig:typeI_coupling}
    \end{subfigure}
    \begin{subfigure}[t]{0.40\textwidth}
        \centering
        \includegraphics[width=\linewidth]{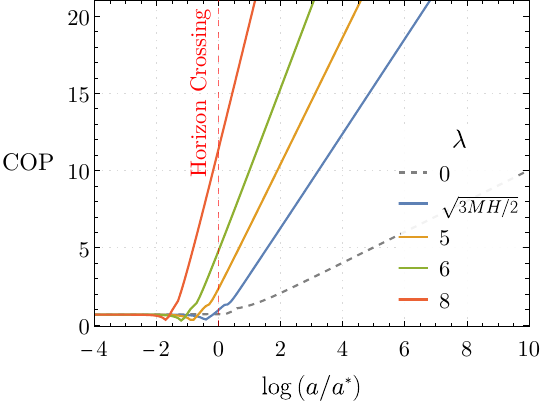} 
        \caption{${\lambda}^2 > 3MH/2$} \label{fig:typeII_coupling}
    \end{subfigure}
    \caption{Time evolution of COP for varying ${\lambda}$ under different regimes of the coupling strength.  Here, we set $m = 0, M = 10, H = 1$ and $k = 1$. The red dashed line shows the timescale of horizon crossing, i.e, $a= k/H$.}
    \label{fig:diff_lambda_COP}
\end{figure}

\par Recall the hierarchy of scales mentioned in Section \ref{sec:model}: $m < \frac{3H}{2} < M$, hence, $mM < \frac{3MH}{2}$, and that we have chosen $m \approx 0$ to make this field a proxy for adiabatic perturbations. This naturally introduces the following ranges of the coupling strength, ${\lambda}^2$, to be considered: (i) $\lambda^2<3MH/2$ and (ii) $\lambda^2>3MH/2$. The weak and strong coupling regimes of ${\lambda}$ exhibit some distinctive features in the time evolution of COP, as listed below (see \cref{fig:COP_generic,fig:diff_lambda_COP}):
\begin{enumerate}

\item[(i)] ${ {\lambda}^2 < 3MH/2}$ : 
COP starts growing linearly after horizon crossing ($a > k/H$). However, after a certain timescale, the slope of the linear growth of COP increases. The transition between the intermediate and late-time growth regions is sharp, and shows a change in the growth rate of COP. The two regions are separated by a bump at $a\equiv a_\text{bump}$ which is a characteristic feature of this case (see \cref{fig:typeI}). As we shall see later on when comparing the COP with the linear entropy, the appearance of the bump in COP signals a decoherence in the model. In the limiting case of ${\lambda}^2\rightarrow 0$, the bump becomes smaller and disappears. As the coupling increases, the timescale of the bump shortens as shown in \cref{fig:typeI_coupling}. In the limit ${\lambda}^2 \rightarrow 3MH/2$, the bump appears at horizon crossing (i.e, $a = k/H$). However, from the perspective of decoherence, what will be important is that the bump signifies an abrupt change in the growth rate of entropy and hence this will be the harbinger of decoherence.
\item[(ii)]  ${\lambda^2 > 3MH/2}$ : 
COP starts growing \textit{before} horizon crossing, i.e, $a<k/H$ (see \cref{fig:typeII}). With increasing coupling $\lambda$, the COP growth starts earlier as shown in \cref{fig:typeII_coupling}.
Unlike the previous case, the linear slope of COP growth does not change at late times; instead, this occurs even before horizon crossing. Moreover, the bump appearing at the change of the COP growth slope is not pronounced in regime (ii) in contrast to regime (i). The change in slope of the COP growth \textit{before} the horizon crossing is simply due to having two effectively heavy fields in de Sitter space,\footnote{Although we have set $m=0$, the coupling term induces a mass for the $\varphi$ field once we integrate out the $\chi$ field.} implying that the characteristic length scale of the system is given by the Compton wavelength of the lighter of the two masses, both of which is smaller than the Hubble radius. Hence, for the bump observed during the change in the slope of complexity growth to serve as a dependable indicator of decoherence, decoherence must take place \textit{prior to} horizon crossing in this scenario.
\end{enumerate}

For this Gaussian system, after diagonalizing the mass matrix and implementing a normal mode decomposition, we find that the eigenvalue $m_l$ is necessarily imaginary (since we set $m=0$) while that corresponding to $m_h$ is always real. However, we observe a significant distinction between the above-mentioned two regimes concerning how the growth of complexity changes with time. The reason behind the change in the growth of complexity can be understood as coming from the effective mass of the $\varphi$ field being light or heavy. The heavy case (corresponding to the so-called ``principal series" representation of de Sitter) is given by regime (ii). In this case, even the light scalar field is fundamentally unstable and decays away \cite{Jatkar:2011ju}. This is why for a large enough coupling ($\lambda$), the heavy mode $m_h > 3H/2$ while for the light mode $|m_l|>3H/2$. In effect, this is what distinguishes regime (ii) from regime (i): In this case, due to both the mass eigenvalues becoming ``heavy'', the change in complexity growth (and, as we shall see later on, decoherence) takes place on sub-horizon scales.\\

What is important, however, for our consideration is that there is necessarily a change in the behaviour of the complexity growth in both regimes. Since the point where this change occurs is what we shall conjecture to be the signal for decoherence in the system, this is the most relevant feature in Fig.~\ref{fig:COP_generic} 
for us. Also, note that for realistic applications to cosmological perturbation theory, coupling strength $\lambda^2 < 3MH/2$ will be the relevant one when $\lambda$ is small enough to have perturbation theory under control. In a Gaussian theory, of course we are free to choose what $\lambda$ is; however, if we think of this term as a leading order interaction in the Lagrangian with higher order terms, then $\lambda$ must be small enough compared to the free Lagrangian to have higher-order interactions parametrically suppressed. \\

Lastly, it's worth noting that the evolution of complexity displays several regions of linear growth, and the slopes within these regions offer insights into their associated Lyapunov exponents \cite{Bhattacharyya:2020art}. We plan to delve into this aspect in a forthcoming study. 
\section{Time Evolution of Linear Entropy} \label{sec:6}

As discussed in section \ref{sec:LinEntropy}, the linear entropy (\(S_l\)) is a good indicator of decoherence introduced by the mixing of two subsystems \(A\) and \(B\). In order to understand the degree to which complexity is sensitive to decoherence in this system, we investigate the time evolution of linear entropy and compare it to that of COP.\\
\begin{figure}[htb!]
    \centering
    \begin{subfigure}[t]{0.40\textwidth}
        \centering
\includegraphics[width=\linewidth]{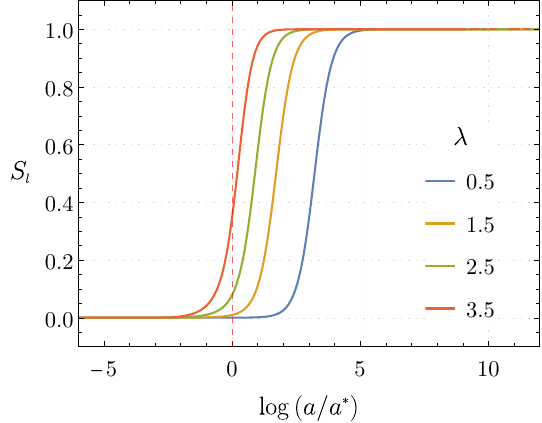} 
        \caption{${\lambda}^2 < 3MH/2$} \label{fig:Sl_typeI_coupling}
    \end{subfigure}
    \begin{subfigure}[t]{0.40\textwidth}
        \centering
        \includegraphics[width=\linewidth]{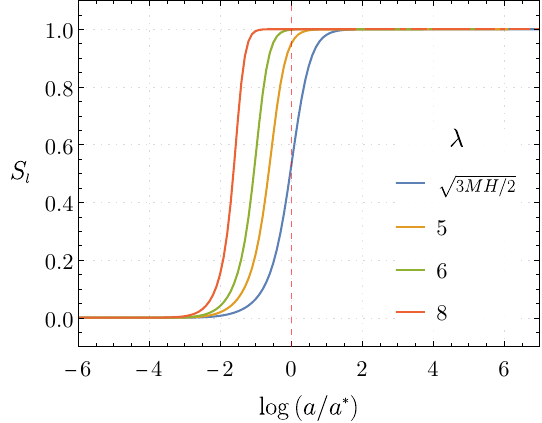} 
        \caption{${\lambda}^2 > 3MH/2$} \label{fig:Sl_typeII_coupling}
    \end{subfigure}
    \caption{Time evolution of the linear entropy $S_l$ for varying ${\lambda}$ under different regimes of the coupling strength. Here, we set $m = 0, M = 10, H = 1$ and $k = 1$. The red dashed line shows the timescale of horizon crossing, i.e, $a= k/H$. 
    }
  \label{fig:Sl_lambda}
\end{figure}

Unlike complexity, the linear entropy does not seem to exhibit characteristically different behavior across the two coupling regimes defined in the previous section. For both regimes, we see that the linear entropy is almost vanishing at early times, undergoes a short period of growth and then saturates, asymptotically approaching unity as the number of \(e\)-folds increases (\cref{fig:Sl_lambda}). Moreover, we observe that the system mixes sooner for greater coupling strength, and tracks with the behaviour of the bump in COP seen in regime (i) (\cref{fig:typeI_coupling}).
\section{Complexity and Decoherence}
\label{sec:decohenrence}
Here, our main objective is to investigate whether complexity can effectively detect decoherence in the curved-space Caldeira-Leggett model. One approach to achieving this goal is to compare the time evolution of the complexity of purification with that of linear entropy and entanglement negativity. Our aim is to identify distinct patterns in complexity behavior when linear entropy and negativity exhibit signs of decoherence.\\

Previously, it was demonstrated in \cite{Bhattacharyya:2022rhm} that in the flat space Caldeira-Leggett model, the saturation timescale of COP aligns with the saturation timescale obtained from linear entropy.\footnote{A comparison with entanglement negativity is also provided in \cref{sec:LogNeg}.} However, in a de Sitter background, COP continues to grow linearly as long as the squeezing parameter keeps increasing, i.e. as long as the modes remain outside the horizon. Consequently, the saturation observed in flat space will not serve as a reliable indicator for detecting decoherence. \\

\begin{figure} [htb!]
   \centering
   \begin{subfigure}[t]{0.32\textwidth}
       \centering
      \includegraphics[width=\linewidth]{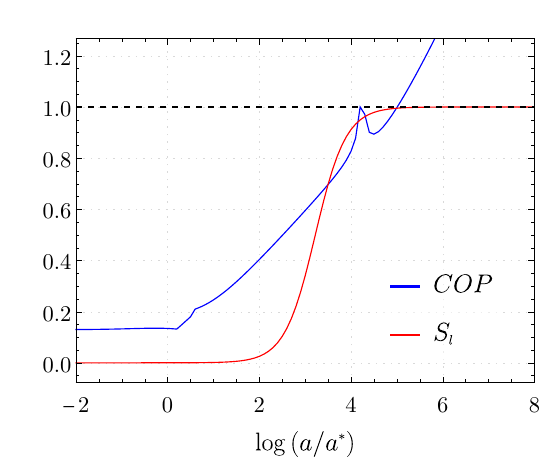} 
        \caption{$\lambda= 0.5$} \label{fig:typeI_compzoom}
    \end{subfigure}
      \begin{subfigure}[t]{0.32\textwidth}
        \centering
        \includegraphics[width=\linewidth]{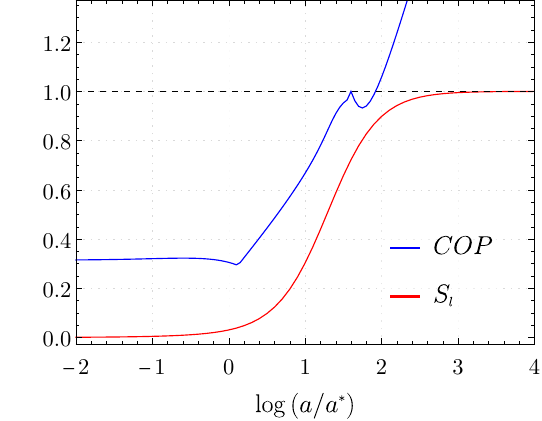}
        \caption{$\lambda= 2.0$} \label{fig:typeII_compzoom}
    \end{subfigure}
    \begin{subfigure}[t]{0.32\textwidth}
    \centering
        \includegraphics[width=\linewidth]{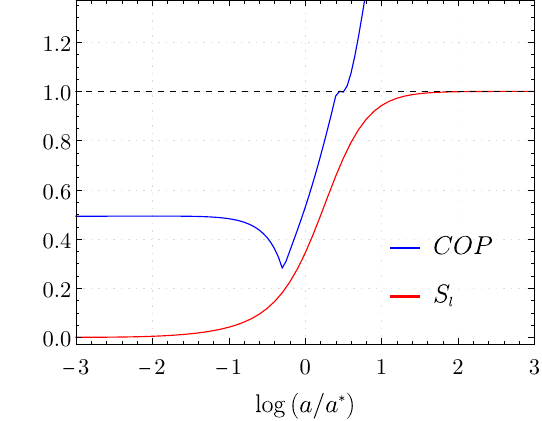}
        \caption{$\lambda=3.5$} \label{fig:typeIII_compzoom}
    \end{subfigure}
    \caption{Time evolution of linear entropy $S_l$, and COP for different values  of coupling strength, where ${\lambda}^2<3MH/2$. Here, we set $m = 0, M = 10, H = 1$ and $k = 1$.  The coupling strength ${\lambda} = 0.5,\, 2.0$ and $3.5,$ in panel a, b and c respectively. COP has been rescaled so that the COP-value at the peak is 1. This is purely for ease of comparison with linear entropy, and does not affect the overall features of COP.}\label{fig:COP_Sl_regime1}
\end{figure}

\noindent

As previously discussed, within the coupling regime ${\lambda}^2 < 3MH/2$, a distinctive feature emerges: the appearance of a bump. Interestingly, the timescale at which this bump manifests align with the saturation timescale of $S_l$. This comparison is illustrated in Fig. \ref{fig:COP_generic}. Since the appearance of the bump varies with the magnitude of the coupling, we delve further into this comparison in Fig. \ref{fig:COP_Sl_regime1}, considering three different values of the coupling within the same regime. Across all these figures, a noteworthy observation is that the decoherence timescale obtained from the saturation timescale of linear entropy concurred with the emergence of the bump regions. 
\begin{figure}[htb!]
\begin{center}
\scalebox{.7}{\includegraphics{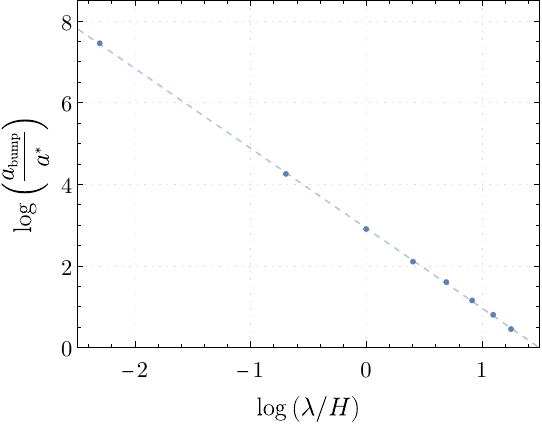} }
 \\
\vspace{-0.1in}
\caption {Dependence of the timescale of the COP bump $a_{\text{bump}}$ on the coupling strength ${\lambda}$ in regime: ${\lambda}^2<{3MH/2}$. Here, we set $m = 0, M = 10, H = 1$ and $k = 1$.}\label{fig:abump} \end{center}
\end{figure}
Furthermore, the plots suggest that the appearance of the bump in COP is a sharper signature for the timescale of mixedness than that of the linear entropy, which saturates asymptotically to unity. Fig. \ref{fig:abump}  displays the location of the bump for various values of the coupling strength $\lambda$. This plot is possible to produce because the bump in COP occurs over a finite window of time. On the other hand, a similar plot for the saturation timescale of linear entropy is only possible if one provides some arbitrarily chosen tolerance specifying the difference between \(S_l\) and unity. \\

\begin{figure}[b!]
   \centering
    \begin{subfigure}[t]{0.40\textwidth}
        \centering
        \includegraphics[width=\linewidth]{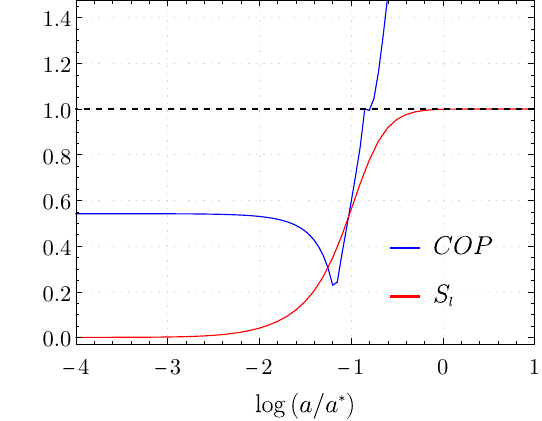} 
           \end{subfigure}
    \caption{Time evolution of linear entropy $S_l$, and COP with ${\lambda}^2>3MH/2$. Here, we set $m = 0, M = 10, H = 1, k = 1$ and ${\lambda} = 6$. }\label{fig:COP_Sl_regime2}
\end{figure}

As mentioned in section \ref{sec:eovCom}, the bump also appears in region (ii), 
although it is not as pronounced as in region (i) and diminishes with increasing coupling. However, even in this regime, we also observe that the saturation timescale for linear entropy coincides with the emergence of the bump, as postulated (see \cref{fig:COP_Sl_regime2}). 

\section{Discussion} \label{sec:8}
Decoherence is a central character in the evolution of inflationary perturbations. This is not just a conceptual issue required to explain the ``classicalization'' of quantum fluctuations, but also affects cosmological observables \cite{Martin:2018lin}. It is typical to track the evolution of the purity (or, equivalently, the linear entropy) of the system modes in order to figure out when they have decohered \cite{Burgess:2022nwu, Gong:2019yyz}. However, it has also been argued that different quantum characteristics of the modes get erased at different points in its evolution history. An example of this is that the quantum discord of states which get decohered ``slowly'', in a de Sitter like expansion, can still remain quite large \cite{Martin:2021znx}. Thus, by itself, it is a reasonable question to ask: when does a mode classicallize?\\

\noindent

All of this gets more complicated due to the effects of curved space. Let us explain this through another example. One typically shows that the density matrix of the system dofs becomes diagonal, in some preferred basis, once the modes exit the Hubble horizon during inflation. However, as is well-known, inflation must be followed by a phase of standard big-bang cosmology, during which the same modes re-enter the horizon. In this case, it is worth asking what guarantees that the state remains diagonal during this post-inflationary evolution. It is true that it is highly unlikely that a quantum state of cosmological perturbations can self-purify by itself; nevertheless, such remarkable phenomenon has been shown to occur for simple Gaussian systems during inflation recently \cite{Colas:2022kfu}. Given this background, it is worthwhile to look for complementary signals which can clearly demonstrate when a pure quantum state becomes mixed during cosmological evolution.\\

\noindent

In this paper, we extend previous work \cite{Bhattacharyya:2022rhm} which showed that complexity can be a good quantifier to show when a system becomes maximally mixed, at least for simple Gaussian systems in a flat background. We extend this study to an accelerating universe to test this hypothesis further. Due to the nature of de Sitter expansion itself, sub-horizon modes which start out in the Bunch-Davies vacuum, get squeezed upon horizon exit. Our metric for computing complexity (or, rather, COP) is such that the effect of this squeezing is to increase it unboundedly. This is in accordance with previous findings \cite{Bhattacharyya:2020rpy} of the squeezing complexity, wherein the complexity only saturated to a finite value due to the assumption of a subsequent radiation phase to inflation. As a qualitative comment, this growth in the complexity for de Sitter space is also in agreement with holographic conjectures \cite{Jorstad:2022mls,Baiguera:2023tpt}, although we leave a more careful comparison for future work.\\

\noindent

Thus, one needs a way to look for features beyond this growth in complexity for it to act as a quantifier for decoherence. We do so by searching for patterns in the evolution of complexity which only reveals features characteristic to decoherence. In the plots for COP, this appears as the timescale on which the rate of growth of the complexity changes since we find later that this correlates very faithfully with the timescale on which linear entropy of the model starts to saturate. Thus, we conjecture, that this is a specific signal for decoherence that can be discerned from the complexity of the system alone. If indeed such a feature (of complexity changing its rate of growth) survives for more complicated, and yet realistic, models of inflation, then this would be an exciting new way to probe for decoherence. For gravity, one has to always contend with non-Gaussian couplings, for which it is very difficult to track the behaviour of the linear entropy. It is precisely for such systems we expect this complementary signal to provide extra hints for the timescale of decoherence.\\

These are early days of application of quantum information theory to cosmological systems. It is, thus, prudent to explore measures such as complexity for realistic cosmological setups and examine their relationship with other well-known quantities such as purity, linear entropy or log-negativity. Not only will these measures be the key to finding observable smoking-gun signals for the quantum origin of inflation, they can even reveal to us which quantum characteristics of inflationary states might retain their coherence at late-times. We hope to report more such exciting results in the future.

\section*{Acknowledgements}
A.B thanks the speakers and participants of the workshop ``Quantum Information in QFT and AdS/CFT-III" organized at IIT Hyderabad between 16-18th September, 2022 and funded by SERB through a Seminar Symposia (SSY) grant (SSY/2022/000446). A.B and S.H. thank the speakers and participants of the workshop ``Quantum Information Theory in Quantum Field Theory and Cosmology" held in 4-9th June, 2023 hosted by Banff International Research Centre at Canada for useful discussions. A.B would like to thank the organizer of the workshop (25-27th July, 2023) ``Integrability, Deformations and Chaos" at OIST, Okinawa and the FISPAC Research Group, Department of Physics, University of Murcia, especially, Jose J. Fernández-Melgarejo, for hospitality during the course of this work. A.B is supported by the Mathematical Research Impact Centric Support Grant (MTR/2021/000490) by the Department of Science and Technology Science and Engineering Research Board (India) and the Relevant Research Project grant (202011BRE03RP06633-BRNS) by the Board Of Research In Nuclear Sciences (BRNS), Department of Atomic Energy (DAE), India. A.B. also acknowledge the associateship program of the Indian Academy of Science, Bengaluru.  S.B. is supported in part by the Higgs Fellowship and by the STFC Consolidated Grant “Particle Physics at the Higgs Centre”. J.L. is supported by the Harry Crossley Research Fellowship.

\appendix
\section{Logarithmic Negativity ($E_\mathcal{N}$)}\label{sec:LogNeg}
When dealing with a mixed state, the entanglement entropy is not a reliable measure of quantum correlation. For instance, the entanglement entropies of subsystems A and B, denoted as $S_A$ and $S_B$, respectively, generally differ for mixed states. In order to capture quantum correlation in such cases, alternative measures have been proposed \cite{Horodecki_2009}. One such measure is the \textit{Entanglement Negativity}. \\

\begin{figure}
   \centering
   \begin{subfigure}[t]{0.40\textwidth}
      \includegraphics[width=\linewidth]{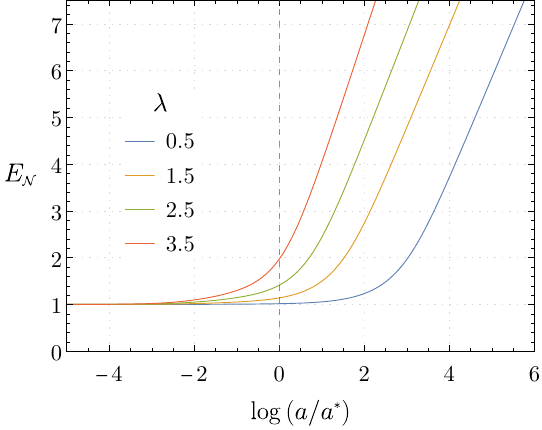} 
      \caption{$\lambda^2 < 3MH/2$} \label{fig:ENtypeI_lambda}
  \end{subfigure}
   \begin{subfigure}[t]{0.40\textwidth}
      \centering
     \includegraphics[width=\linewidth]{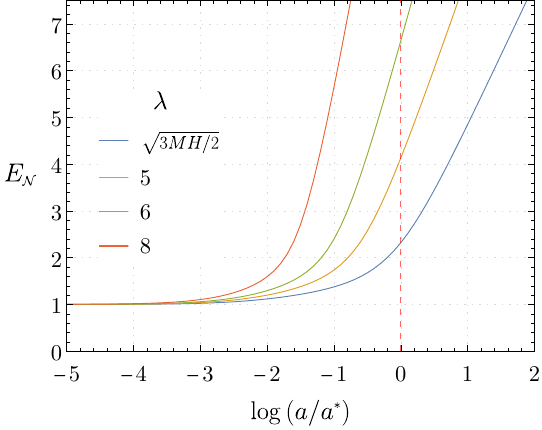} 
     \caption{$\lambda^2 > 3MH/2$} \label{fig:ENtypeII_lambda}
   \end{subfigure}

   \caption{The time evolution of logarithmic negativity $E_\mathcal{N}$ for varying $\lambda$ under different regimes of the coupling strength.  Here, we set $m = 0, M = 10$, $H=1$ and $k = 1$. The red dashed line shows the timescale of horizon crossing, i.e, $a= k/H$. }\label{fig:logneg}
\end{figure}
The entanglement negativity is defined by taking the trace norm of the partial transpose of the density matrix. Consider a bipartite state $\hat{\rho}_{AB}$ consisting of subsystems A and B. The partial transposition of $\hat{\rho}_{AB}$ with respect to subsystem B, expressed in a local orthonormal basis as $\hat{\rho} = \sum \rho_{ij,kl}\ket{i}\bra{j} \otimes \ket{k}\bra{l}$, is defined as:
\begin{align}
\hat{\rho}^{T_B} := \sum_{i,j,k,l} \rho_{ij,kl}\ket{i}\bra{j} \otimes \ket{l}\bra{k}\,.
\end{align}
It is important to note that the spectrum of the partial transpose of the density matrix remains unchanged regardless of the chosen basis or whether the partial transposition is performed over subsystem A or B. The positivity of the partial transpose of a state serves as a necessary condition for separability of the density matrix \cite{Peres_1996}. This separability criterion motivates the consideration of entanglement negativity, which is related to the absolute value of the sum of the negative eigenvalues of $\hat{\rho}^{T_B}$ \cite{Vidal_2002, Eisert_1999, plenio2006introduction}. Specifically, the entanglement negativity $\mathcal{N}(\hat{\rho})$ is defined as:
\begin{align}
\mathcal{N}(\hat{\rho}) := \frac{1}{2}\left(|\hat{\rho}^{T_B}|-1\right)\,,\label{negativity}
\end{align}
where $|A|\equiv \rm{Tr}{\sqrt{A^\dagger A}}$ represents the trace norm. It can be shown that $\mathcal{N}(\rho)$ is zero for unentangled states, i.e., for separable density matrices $\hat{\rho}$. Another relevant quantity, known as the \textit{Logarithmic Negativity}, denoted as $E_{\mathcal{N}}$, is defined as follows:
\begin{align}
E_{\mathcal{N}}(\hat{\rho}) := \log_2{\left|\hat{\rho}^{T_B}\right|} = \log_2{\left(2\mathcal{N}(\hat{\rho})-1\right)}\,.
\end{align}
These measures have gained attention in recent times in the study of field theory and quantum many-body systems \cite{Calabrese:2012ew,Calabrese:2014yza,Calabrese:2012nk,Calabrese:2013mi,Alba_2013,PhysRevB.90.064401,Ruggiero_2016,Ruggieroa_2016,Blondeau_Fournier_2016,Eisler_2014,Coser_2014,Hoogeveen_2015,PhysRevB.92.075109,PhysRevA.88.042319,PhysRevA.88.042318,Wen_2016, Wena_2016,Kudler_Flam_2019,Verstraete_2001}. \\

As we are dealing with a Gaussian state, we may calculate the logarithmic negativity from the covariance matrix \cite{Adesso2005}. The partial transpose of the reduced density matrix $\rho$ corresponds to a transformation of the covariance matrix $\Sigma \rightarrow \tilde{\Sigma}$ which flips the sign of the determinant of its off-diagonal block: $\det\Sigma^{(\ph\chi)} \rightarrow - \det\Sigma^{(\ph\chi)}$. The logarithmic negativity is then given by:
\begin{equation}
    E_{\mathcal{N}} = \text{max}[0,-\log_2(\tilde{\nu}_-)]\,,
\end{equation}
Where $\tilde{\nu}_-$ is one of the symplectic eigenvalues of $\tilde{\Sigma}$ given by:
\begin{equation}
    \tilde{\nu}_{\mp} = \sqrt{\frac{\Delta(\tilde{\Sigma}) \mp \sqrt{(\Delta(\tilde{\Sigma}))^2 - 4\det\Sigma}}{2}}\,,
\end{equation}
where
\begin{equation}
    \Delta(\tilde{\Sigma}) \equiv \det \Sigma^{(\ph\ph)} + \det \Sigma^{(\chi\chi)} - 2\det \Sigma^{(\ph\chi)}\,.
\end{equation}
\begin{figure}
    \centering
    \begin{subfigure}[t]{0.40\textwidth}
        \includegraphics[width=\linewidth]{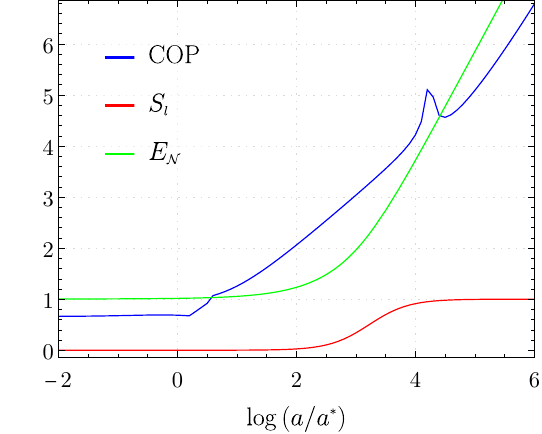} 
        \caption{$\lambda^2 < 3MH/2$} \label{fig:typeI_comp_EN}
    \end{subfigure}
    \begin{subfigure}[t]{0.40\textwidth}
        \includegraphics[width=\linewidth]{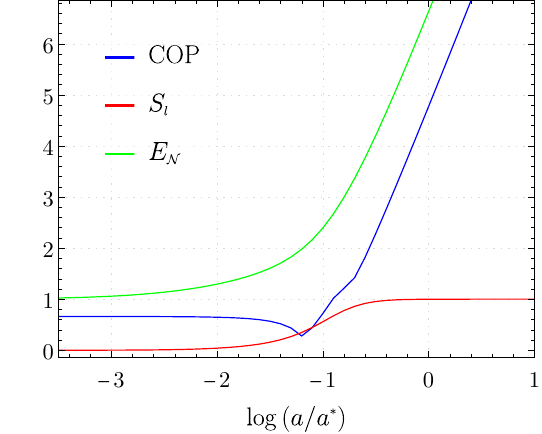} 
        \caption{$ \lambda^2 > 3MH/2$} \label{fig:typeII_comp_EN}
    \end{subfigure}
   
    \caption{The time evolution of linear entropy $S_l$, logarithmic negativity $E_{\mathcal{N}}$, and COP under different regimes of the coupling strength. Here, we set $m = 0, M = 10, H = 1$ and $k = 1$.  The coupling strength ${\lambda} = 0.5$ and $6$ in the left and the right panel respectively.
    }\label{fig:allthree}
\end{figure}
\section{Time Evolution of Logarithmic Negativity} \label{app:neg}
For the sake of comparison between logarithmic negativity (\(E_{\mathcal{N}}\)), linear entropy ($S_l$) and complexity (COP), we discuss the effect of the coupling \(\lambda\) on the time evolution of negativity.\\

The logarithmic negativity seems to be insensitive to the two regimes defined in \cref{sec:eovCom}. Regardless of the general range of values which the coupling \({\lambda}\) falls into, the overall shape of the \(E_{\mathcal{N}}\) curve appears similar: at early times the negativity is close to 1, then undergoes a period of rapid growth, and seems to approach a linearly increasing trajectory as \(\ln(a/a^*) \rightarrow \infty\) (see \cref{fig:logneg}). Notably, for both regimes, increasing \(\lambda\) has the effect of decreasing the timescale at which the logarithmic negativity starts to grow. This is shown in \cref{fig:allthree}. This is to be expected - a larger coupling \(\lambda\) corresponds to a greater degree of ``mixing'' between the fields \(\ph\) and \(\chi\), and the logarithmic negativity should detect this as it has been seen to be a reliable measure of quantum correlation \cite{Vidal_2002}.

\FloatBarrier
\bibliographystyle{JHEP}
\bibliography{biblio}

\end{document}